# Measures of Cluster Informativeness for Medical Evidence Aggregation and Dissemination

Michael Segundo Ortiz[1], Sam Bubnovich[1], Mengqian Wang[1] Kazuhiro Seki, PhD[2], Javed Mostafa, PhD[1]
[1]University of North Carolina, Chapel Hill, NC, USA
[2]Konan University, Kobe, Japan

**Abstract**

*The largest collection of medical evidence in the world is PubMed. However, the significant barrier in accessing and extracting information is information organization. A factor that contributes towards this barrier is managing medical controlled vocabularies that allow us to systematically and consistently organize, index, and search biomedical literature. Additionally, from users' perspective, to ultimately improve access, visualization is likely to play a powerful role. There is a strong link between information organization and information visualization, as many powerful visualizations depend on clustering methods. To improve visualization, therefore, one has to develop concrete and scalable measures for vocabularies used in indexing and their impact on document clustering. The focus of this study is on the development and evaluation of clustering methods. The paper concludes with demonstration of downstream network visualizations and their impact on discovering potentially valuable and latent genetic and molecular associations.*

**Introduction**

The most basic clustering problem involves taking a set of items and splitting them into smaller groups in such a way that the items within each group are enough alike that one can ignore the slight differences between the items. As the problem grows in complexity, the tradeoff is a balance between accuracy, computational efficiency, and most importantly, informativeness. In the biomedical domain, operationalizing these interwoven elements is further complicated by the rate of publication (Figure 1).

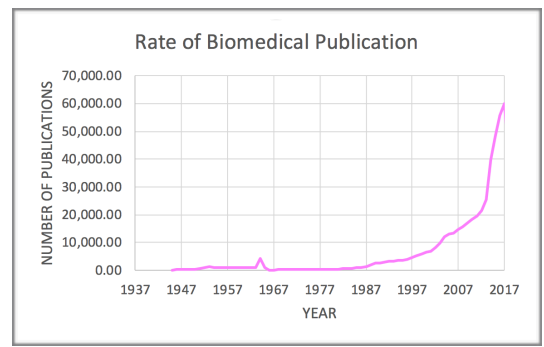

**Figure 1**. PubMed indexed publications with search term "biomedical" in any field (biomedical[All Fields]).

The formation of clusters is used primarily to facilitate the retrieval of similarly grouped and relevant concepts. However, an essential problem is the evaluation or measurement of clustering solutions in order to objectively assess the informativeness while exploring the literature at various stages in a biomedical pipeline. In order to accomplish this task, the information must first be organized. Thus, the science of information relies heavily on the early stages of information representation. Informaticians may process documents by vectorizing a corpus of interest and represent each term or document as a point in vector space[1]. This generates a term-document matrix that maps all terms to a particular document id. From the raw matrix, weighting schemes are applied that quantify the representativeness of a feature. Term-frequency and inverse document frequency ($tf \times idf$) is the most widely used and performs well[2,3]. The intuition behind this weighting scheme is to identify features within a body of text that have a balance between high intra-document frequency, and low inter-document frequency. The information can then be rearranged into a correlation or document by document matrix which forms the basis for similarity measures and subsequent clustering algorithms for association discovery[4]. To reduce noise, computational load, and account for co-occurrence, truncated singular value decomposition, also known as latent semantic analysis or LSA, can be applied to reduce the rank of the matrix and generate a dimensionally reduced information space[5].

Clustering follows a simple paradigm and is an optimization problem at its core. Fundamentally, it creates a representation where each individual term or document is instead represented by the "typical" or the average characteristics of its class. The variability in a single cluster $c$ is affected by adjusting various parameters and described as the sum of the squared distances of each point from the centroid (mean) and is a Euclidean or "as the crow files" distance metric:

$$variability\,(c) = \sum distance\,(mean\,(c), e)^2$$

We can extend this basic formula of single cluster $c$ variability, to measure the level of dissimilarity among a group of clusters $C$ which is the sum of all observed single cluster variabilities:

$$dissimilarity\,(C) = \sum variability\,(c)$$

Following this logic, one might assume that the most reasonable approach toward a clustering solution would be to minimize the objective function $C$. In other words, to reduce the dissimilarity among a group of clusters. However, by simply having the same number of clusters as observations, and assigning each observation to its own cluster, variability would be zero, dissimilarity would be zero, and a theoretically perfect solution can be found with absolutely no aggregation of evidence or informativeness. This is precisely why constraints, in the form of parameter tuning, are utilized. There are various parameters that can be adjusted to optimize a downstream clustering solution. Upstream, one can adjust the threshold for document frequency and allow only terms satisfying the threshold to be weighted in the matrix representation of a corpus. One can also adjust a threshold for the ranking of $tf \times idf$ weights for each document, applying a strict cutoff to the matrix to subsequently cluster only the top ranked terms. Once various tuning for feature representation has been completed, constraints on the clusters themselves can be applied, such as the number of clusters. It is apparent that optimal parameter tuning to extract informativeness is qualitative and herein lies a motivation for this work. The quantitative evaluation of various tuning adjustments, combinations in parameter space, and the direct relationship to cluster informativeness and evidence aggregation is what we seek to demonstrate.

The classic evaluation measures for information retrieval applications are precision and recall. Imagine a query and a ranked list of retrieved documents, if every document in the retrieved list is relevant to the query, then precision is perfect, however if some relevant documents are not retrieved in the list, then recall is imperfect. The harmonic mean of these two measurements is known as the F-measure[6]. The higher the F-measure, the better overall performance and quality of the retrieval model and thus a balance between precision and recall is rewarded. For visual information retrieval, specifically clustering, evaluation measures such as homogeneity, completeness, and v-measure are utilized. Homogeneity is much like precision; for an unsupervised cluster, homogeneity evaluates the relatedness of intra-cluster components. As an example, if a sub-topic cluster for breast cancer SNPs, contains only documents with high similarity to breast cancer SNPs and no documents related to breast cancer treatments, then this unsupervised cluster would be perfectly homogenous. However, if there are similar documents for breast cancer SNPs also found in a different cluster, for example, in breast cancer immunotherapy, then the breast cancer SNP cluster is homogenous but not complete. Thus, the completeness metric is intuitively similar to recall. The harmonic mean of homogeneity and completeness is known as the v-measure which measures the overall "validity" of the clustering solution[7], or how "close" a clustering solution can get to a gold standard.

**Methods**

Dataset preparation

We downloaded and indexed the MEDLINE database from the National Library of Medicine servers onto a laboratory server and retrieved articles related to breast cancer with the query, "breast neoplasms"[MeSH Terms], which resulted in 264,337 scientific abstracts. Articles annotated with major MeSH terms (descriptors or qualifiers) below "breast neoplasms" (direct children) in the MeSH hierarchy were filtered and the four most frequent were used as our first experimental dataset. This resulted in a total of 4 major MeSH terms distributed across 16,305 abstracts which served as class labels for downstream clustering and evaluation. The PubMed identifiers were extracted from the abstracts, cross-referenced to PubMed Central, and used to create a corpus of 1,400 open access full-text articles (Figure 2). Not every abstract had an open access full-text article as of 7-16-2018. The two experimental conditions for measuring cluster informativeness were 1,400 abstracts versus 1,400 full-text articles.

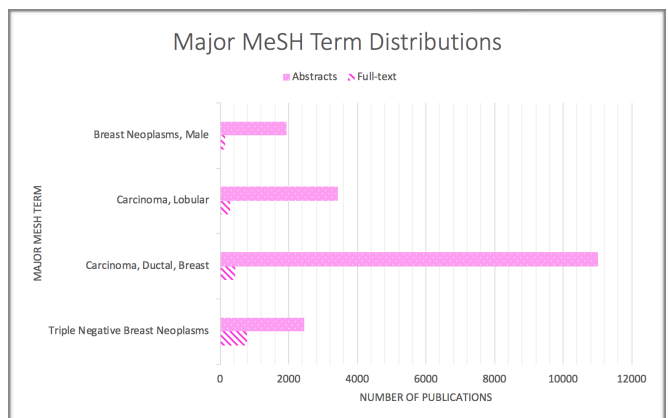

**Figure 2**. MeSH distributions.

Corpora processing

Each experimental dataset — abstract and full-text — were processed by the same protocol that follows. The corpora were vectorized into a term-document matrix and $tf \times idf$ weights for each feature were generated by the following formula:

$$weight_{term} = t_c \times \log\left(\frac{N}{n_t}\right)$$

where $t_c$ is the number of occurrences for a particular term in each indexed document, $N$ is the total number of documents, and $n_t$ is the number of documents the term is occurring in. We then removed features from the matrices that had $n_t = 1$. The rationale for feature ablation by this constraint is that natural language and random text both exhibit an inverse power-law distribution whereby a significant proportion of features occur so infrequently that they offer little information and confound the representation[8]. By ablating features that only occurred in one document, we were able to reduce noise and computational load by approximately 50%.

Clustering

There are four parameters that dictate a clustering solution and as a consequence, informativeness. $D$, $R$, $N$ and $K$. We can further adjust document frequency threshold with $D$. Recall that in processing the corpora, we applied a constraint of $n_t = 1$, meaning that a term must occur in more than one document across the entire collection to be included in our analysis. By implementing the control knob $D$, we were able to explore how clustering solutions were affected by varying degrees of document frequency for terms in combination with other parameters. The ranking of weights across term vectors was also parameterized with the control knob $R$ which allowed us to set cutoffs for the vectors and only cluster associations that met the minimum weight criteria. The $N$ parameter allows for the application of eigenvalue analysis to dimensionally reduce the matrix by preserving nonzero term vectors — eigenvectors — and similarity structure[9]. Lastly, given the underlying MeSH structure of our experimental datasets, the additional $K$ parameter for the K-means clustering algorithm should logically be $K = 4$, i.e., Four clusters that correspond to the four major MeSH terms. However, as part of our analysis, we wished to examine how the combination of the parameters affect the natural clustering and informativeness of breast cancer sub-topics considering that gold standard MeSH annotation is conducted by human annotators at the National Library of Medicine. Thus, we developed an experimental matrix with lower and upper bounds for each parameter with random iteration to evaluate each combination for both experimental conditions (Table 1).

| Parameter | Lower & Upper Bounds |
|---|---|
| $D$ (document frequency) | $0.1\% - 1.0\%$ |
| $R$ (tf x idf rank cutoff) | $5 - 14$ |
| $N$ (number of dimensions) | $1 - 20$ |
| $K$ (number of clusters) | $2 - 20$ |

**Table 1**. Experimental matrix for PubMed abstracts and PubMed Central full-text, random iteration over all parameter combinations.

Breast cancer dictionary

Without genetic and molecular context, the evaluated clusters offer no informativeness for evidence aggregation and dissemination, with respect to breast cancer. Thus, the final dimension of our analysis was focused on automating the extraction of genetic and molecular content from evaluated clusters. We queried the NCBI Gene database for known or predicted genes defined by nucleotide sequence or map position with the query "*Breast Cancer AND "Homo sapiens"[porgn:__txid9606] tax_id*". This search resulted in 3,766 genes as of 7-23-2018. We extracted *Gene Symbol, Aliases,* and *Molecular Description* for each gene into a dictionary with the following data structure:

*{{Gene Symbol : [Alias$_1$, Alias$_2$,...,Alias$_n$]} : Molecular Description}*

With this dictionary, we probed the clusters across each experimental condition for a unique network of genes and molecular descriptions. The challenge for this automation is that of quantifying unique genes and molecular descriptions that are highly representative for a cluster. This required a normalization step. The identification of information in each cluster is a trivial task and requires only conditional logic with a programming language of choice. However, in order to quantify unique genetic and molecular units of information in a cluster we first had to establish a global average for all genes or molecular descriptors across clusters where $G_k, M_k$ are the number of occurrences for a gene or molecular descriptor in a cluster, $G_{global}, M_{global}$ are the number of occurrences for a gene or molecular descriptor across all clusters, $D_k$ is the number of documents in a cluster, and $D_{total}$ is the total number of documents across all clusters. We then ordered relative informativeness by the most highly weighted units of genetic or molecular information in each cluster and hypothesized that the highest relative units of information would characterize the cluster and reveal networks of genes or molecular pathways that are associated:

$$Gene, Molecular_{relative-expression} = G_k, M_k - \frac{G_{global}, M_{global} \times D_k}{D_{total}}$$

**Results**

An essential starting point for our analysis was to decide upon default parameters in order to have a baseline of performance with respect to the evaluation of cluster informativeness against the MeSH gold standard. Based on previous work[4], we chose $R = 5$, and $D = 0.5\%$. Choosing the default dimensionality $N$, for truncated singular value decomposition, was experimentally chosen by randomly sampling 1,000 abstracts from the original 16,305 abstracts and iterating through the experimental matrix. We observed that $N = 15$ resulted in the highest v-measure. We then analyzed our datasets — PubMed abstracts and PubMed Central full-text — based on the default parameters for $R, D, N$ and plotted v-measure as a function of $K$ in order to observe the optimal number of clusters as a baseline (Figure 3). We followed up with a parameter sweep to find the optimum combination of $D, R, N$ and $K$ with respect to completeness, homogeneity, and v-measure (Table 2 & 3). The informativeness of the most performant clustering solution in each experimental condition was examined using our breast cancer dictionary as a probe, and derived algorithmically by the gene and molecular description weighting scheme. We observed that unique networks of genes and descriptions emerged from the clusters (Figures 4a-5b).

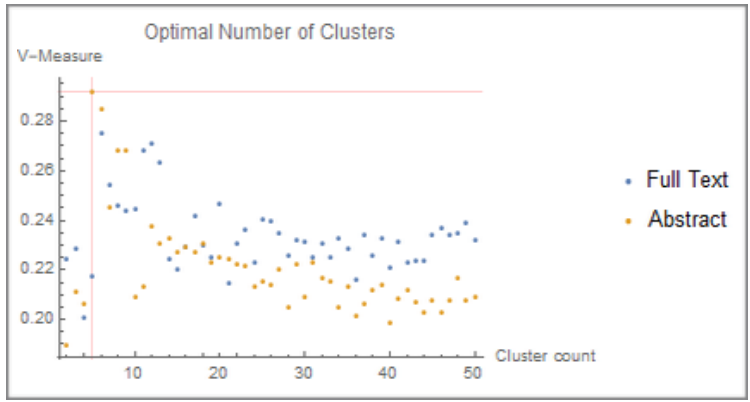

**Figure 3**. V-Measure as a function of K, optimal number of clusters K = 4, with V-Measure = 0.29.

| PubMed Abstracts | | | | | | |
|---|---|---|---|---|---|---|
| $D$ | $R$ | $N$ | $K$ | $Completeness$ | $Homogeneity$ | $V-Measure$ |
| 0.6 | 6 | 13 | 4 | 0.337 | 0.327 | 0.332 |
| 0.6 | 6 | 12 | 4 | 0.336 | 0.327 | 0.331 |
| 0.8 | 8 | 11 | 4 | 0.335 | 0.325 | 0.330 |
| 0.8 | 6 | 10 | 4 | 0.330 | 0.314 | 0.322 |
| 1 | 5 | 16 | 5 | 0.360 | 0.285 | 0.318 |

**Table 2**. Parameter sweep, PubMed abstracts, top five results, ranked by v-measure.

| PubMed Central Full-Text | | | | | | |
|---|---|---|---|---|---|---|
| $D$ | $R$ | $N$ | $K$ | $Completeness$ | $Homogeneity$ | $V-Measure$ |
| 0.2 | 6 | 9 | 4 | 0.361 | 0.325 | 0.342 |
| 0.5 | 6 | 15 | 5 | 0.374 | 0.309 | 0.338 |
| 0.3 | 8 | 8 | 6 | 0.403 | 0.284 | 0.333 |
| 0.5 | 6 | 10 | 6 | 0.384 | 0.294 | 0.333 |
| 0.7 | 5 | 15 | 7 | 0.402 | 0.282 | 0.332 |

**Table 3**. Parameter sweep, PubMed Central full-text, top five results, ranked by v-measure.

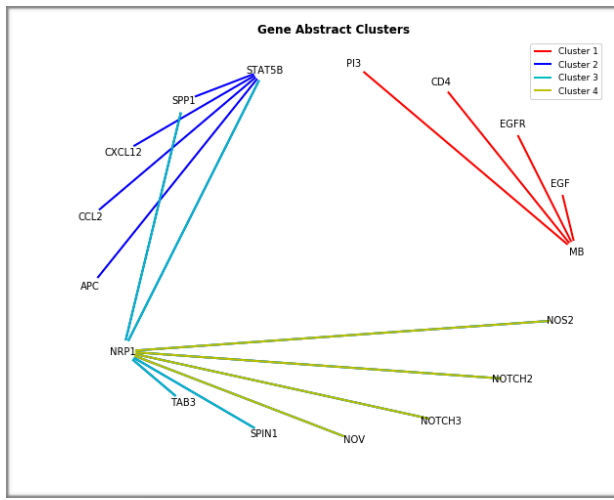

**Figure 4a**. Abstracts, network of relative gene symbol or alias weights, top five ranking for each cluster.

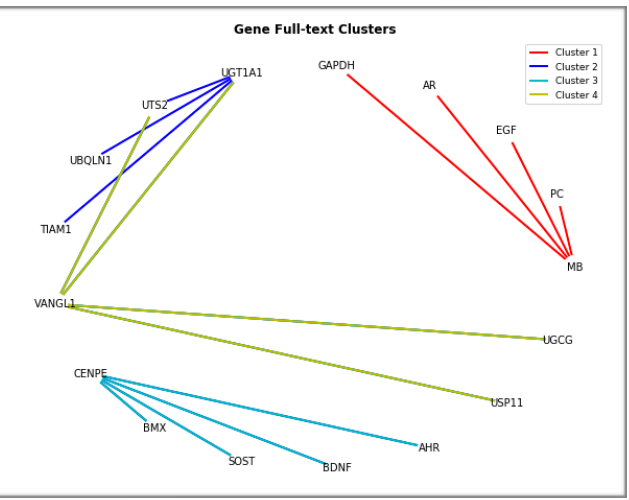

**Figure 4b**. Full-text, network of relative gene symbol or alias weights, top five ranking for each cluster.

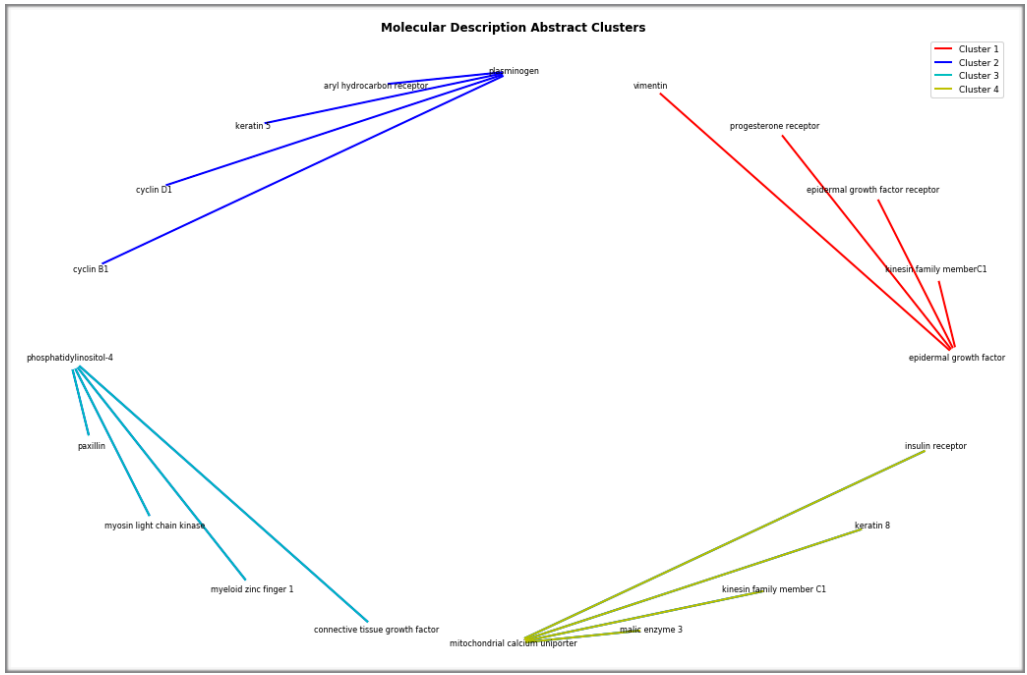

**Figure 5a**. Abstracts, network of relative molecular description weights, top five ranking for each cluster.

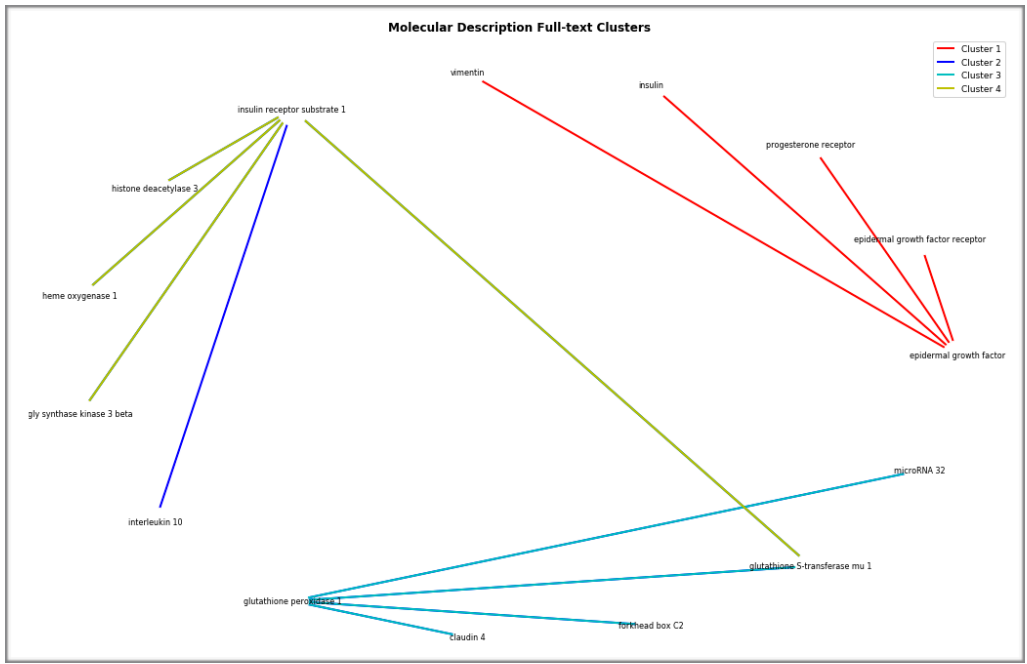

**Figure 5b**. Full-text, network of relative molecular description weights, top five ranking for each cluster.

**Discussion**

The evaluation methods developed in this study attempted to assess the quality of clustering by utilizing MeSH as the "anchor" or "gold standard" vocabulary set. However, we realize that inter-indexer consistency and MeSH creation can include inconsistencies, as it is an innately and exclusively human activity. Overall, the aim was to demonstrate how close the automated indexing and clustering approach was to the one which could be achieved by solely relying on MeSH. Our results and observations indicate that we are one-third of the way to accomplishing this task. A v-measure of 33% for abstract clustering and 34% for full-text clustering may not be performant. Although, the result, as compared to an exclusively human-based approach seem less than superior, in our consideration the attempt to advance automated indexing and clusteirng, and developing a logical way to measure the outcomes are some of the key contributions of this study. Moreover, the nodes and edges generated in our networks show that in some instances the clusters are not visually homogenous. From an evaluation perspective one may reason that this needs improvement, however, when we leveraged the breast cancer dictionary and gene/molecular weighting scheme, network connectivity emerges across clusters. Thus, homogeneity may have limitations with respect to latent information and associations. It is important to note that our approach cannot be seen in the same light as gene regulatory networks would be seen to an experimentalist or clinician. To clarify, The clustering and subsequent network connectivity of genes and molecular descriptions does not imply that they are directly interacting with each other at the cellular level. Perhaps novel associations will in-fact be observed that encourage experimentation and validation[10–12], however the associations themselves are information-theoretics with mathematical underpinnings.

**Conclusion**

Nearly 60,000 publications were indexed in PubMed in 2017 relating to biomedicine; 164 publications per day. As of July 2018, 40,000 publications have already been indexed with five months left in the year. A sequential list of documents is the current model of retrieval for biomedical literature. Only a small portion of the relevant documents are ever explored. This element of human-computer interaction was not investigated in this work but rather laying the foundation toward the goal of evolving from the "please come to page two" model. The transformation of a sequential concept space to a flat concept space with clusters and networks of information will be the future model of information retrieval in medicine because the connected nature of human biology and medical advancement demands equally connected information spaces that aid in the retrieval of relevantly associated information. Network evaluation, approximations, and heuristics will be helpful in such visual information retrieval systems in medicine because clusters can be explored at a deeper level whereby the semantic relationship between units of information within one cluster and across clusters can be analyzed as opposed to a more foundational approach where units of information are analyzed to derive which cluster they naturally belong to.

**Acknowledgements**

The NIH-NLM T15 grant 5T15LM012500-02 provided support for the graduate student research assitants.